\documentclass[aps,pre,amsmath,amssymb,twocolumn,showpacs,byrevtex,superscriptaddress]{revtex4}
\usepackage{graphicx}
\begin{document}

\title{Model for reversible colloidal gelation}

\author{E. Zaccarelli}
\affiliation{
Dipartimento di Fisica and INFM-CRS SOFT, Universit\`a di Roma `La
Sapienza', P.le A. Moro~2, I-00185, Roma, Italy}

\author{S.~V. Buldyrev}
\affiliation{
Yeshiva University,  Department of
Physics, 500 W 185th Street New York, NY 10033, USA}

\author{E. La Nave}
\affiliation{
Dipartimento di Fisica and INFM-CRS SOFT, Universit\`a di Roma `La
Sapienza', P.le A. Moro~2, I-00185, Roma, Italy}

\author{A.~J. Moreno}
\affiliation{
Dipartimento di Fisica and INFM-CRS SMC, Universit\`a di Roma `La
Sapienza', P.le A. Moro~2, I-00185, Roma, Italy}

\author{I. Saika-Voivod}
\affiliation{Dipartimento di Fisica, 
Universit\`a di Roma `La Sapienza', P.le Aldo
Moro~2, I-00185, Roma, Italy}

\author{F. Sciortino}
\affiliation{
Dipartimento di Fisica and INFM-CRS SOFT, Universit\`a di Roma `La
Sapienza', P.le A. Moro~2, I-00185, Roma, Italy}

\author{P. Tartaglia}
\affiliation{
Dipartimento di Fisica and INFM-CRS SMC, Universit\`a di Roma `La
Sapienza', P.le A. Moro~2, I-00185, Roma, Italy}

\date{\today}

\begin{abstract}

We report a numerical study, covering a wide range of packing fraction
$\phi$ and temperature $T$, for a system of particles interacting via
a square well potential supplemented by an additional constraint on
the maximum number $n_{\rm max}$ of bonded interactions.  We show that
when $n_{\rm max}<6$, the liquid-gas coexistence region shrinks,
giving access to regions of low $\phi$ where dynamics can be followed
down to low $T$ without an intervening phase separation.  We
characterize these arrested states at low densities (gel states) in
terms of structure and dynamical slowing down,
pointing out features
which are very different from the standard glassy states observed at
high $\phi$ values.
\end{abstract}

\pacs{82.70.Gg, 82.70.Dd, 61.20.Lc}

\maketitle

Extensive investigations have recently focused on slow dynamics in
colloidal systems, triggered by the experimental discovery of multiple
mechanisms leading to a disordered arrested state.  At high packing
fraction $\phi$, arrest takes place via a glass transition process,
which can be driven by jamming, as in hard-sphere (HS) systems, or by
attractive bonding between colloidal particles, generating the
so-called ``attractive glass''\cite{pham,eckert02,natmat}. At low
$\phi$, arrest takes place via particle clustering\cite{prlweitz} and
formation of an amorphous state of matter --- named a gel --- able to
support weak stresses\cite{lekker-exp-gel}. Clustering and gelation
have no counterpart in atomic or molecular systems, being induced by
the presence of attractive interactions between particles, with a
range of interaction much smaller than the particle size.

\begin{figure}
\includegraphics[width=.35\textwidth]{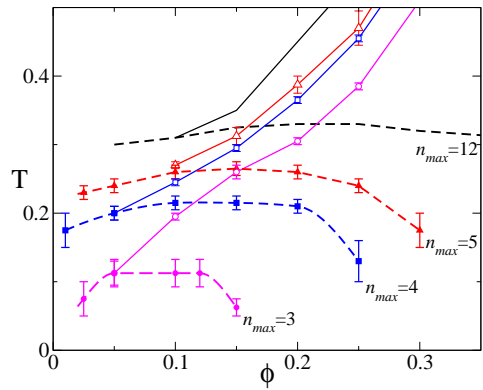}
\caption{
Spinodal (dashed) and percolation (full) loci (with reported error
bars) for $n_{\rm max}=3$ (circles), $4$ (squares), $5$ (triangles)
and 12 (no symbols). Lines are guide to the eye.  }
\label{fig:perc_spin}
\end{figure}

Recently, several studies have addressed the question of the routes to
the gel state in colloidal
systems~\cite{fuchs,langmuir,kroy,yukawa,sator,puertas,delgado}.  In
most models proposed so far, phase separation or microphase separation
provides the initial step of the gelation process. For short-ranged
attractive colloids, at low $\phi$ and temperature $T$, the phase
diagram is characterized by a flat phase-coexistence curve (e.g. see
Fig.~\ref{fig:perc_spin}). A quench inside the two-phase region
induces, via spinodal decomposition, a separation into colloid rich
(liquid) and colloid poor (gas) phases.  However, in appropriate
conditions, dynamical arrest in the denser region, of the attractive
glass type, intervenes by freezing the pattern generated during the
coarsening process. The time to arrest and the structure of the formed
gel depends on $\phi$ and on the interaction strength.  The connection
between gelation and phase
separation~\cite{bos,jackle,heyes1,soga,capri} is supported by the
experimental observation of a peak in the scattered intensity at small
wavevectors\cite{prlweitz,zukoski}.  In the case of phase separation
induced gelation, the slowing down of the dynamics associated with the
formation of an arrested state can not be continuously followed
through equilibrium states, since gelation takes place only after a
quench inside the coexistence region. This is different from the
slowing down of the dynamics on approaching the glass transition at
high densities, for which the $\phi$ and $T$ dependence of the slow
dynamics can be characterized in terms of (metastable) equilibrium
states in a reproducible and reversible way.

In principle, gelation could also occur in the absence of any phase
separation, if the liquid-gas separation is suppressed or if the gel
line is encountered before the
liquid-gas coexistence locus\cite{fuchs}, preventing phase separation.
In these cases, the approach to the gel state could be in principle
followed in equilibrium and reversibly.  Some experimental
groups\cite{zukoski} do indeed favor this interpretation, explaining
the increase in the scattered intensity as intrinsically due to the
inhomogeneities associated with the gel structure, posing a challenge
to theoreticians to develop a model where arrest at low $\phi$ is
observed in the absence of a phase separation mechanism and in which
reversible physical gelation can be studied in equilibrium.
 
In this Letter we introduce and numerically study a simple model
showing that dynamical arrest at low $\phi$ (as low as $0.2$) can
indeed be generated in the absence of phase separation. We follow,
via extensive simulations, the evolution, in equilibrium, of the
density autocorrelation functions to point out the differences in the
dynamics of a gel and of a glass, and to provide a way of
discriminating between the two phenomena.  

We study a modification of the short-range square well (SW) potential,
by adding a constraint which limits the maximum number of bonds
$n_{\rm max}$ that can be formed by the particles\cite{pine}. Without
the constraint, each particle could in principle interact
simultaneously with the maximum number of neighbours allowed by
geometrical packing ($n_{\rm max}=12$).  The constraint switches off
the attractive well potential when any of the two interacting
particles has already $n_{\rm max}$ bonded neighbors. In this way, the
energy difference between particles located in the interior of an
aggregate and particles at the aggregate surface can be decreased and
even suppressed. As a result, the surface tension
decreases\cite{hill}, the driving force for phase separation is
significantly reduced and open structures are favored. 
We find that for $n_{\rm max}<6$, the system can access the $\phi$
region usually dominated by phase separation and experience a
dynamical slowing down by several orders of magnitude, thus entering
the gel regime. Remarkably, the system undergoes the fluid-gel
transition {\it in equilibrium}, so that the process is fully
reversible. 

We perform event-driven molecular dynamics (MD)
simulations of particles interacting via a maximum valency
model\cite{speedy}, i.e. a SW potential, where particles can form a
maximum number of bonds $n_{\rm max}$. We fix the depth $u_0=1$, and
the width $\Delta/(\sigma+\Delta)=0.03$, with $\sigma=1$ being the
particle hard-core diameter. The square well form of the potential
unambiguously defines bonded particles when particle centers lie
within a distance $\sigma < r < \sigma + \Delta$ from each other. When
a particle is already bonded to $n_{\rm max}$ neighbors, the well
interaction with other particles is switched off leaving only the HS
potential.  We simulate a large system containing $N=10^4$ particles
of mass $m=1$ to minimize finite size effects.  Temperature is
measured in units of $u_0$, time $t$ in units of $\sigma
(m/u_0)^{1/2}$.  For all simulated state points, we first equilibrate
the system at constant $T$ until the potential energy and pressure $P$
of the system have reached a steady state, and particles have diffused
several $\sigma$'s on average. Then, a production run is performed at
constant energy. An average over typically 100 different realizations
is done to gather statistics. We focus on the cases $n_{\rm max}=3$,
$4$ and $5$, since $n_{\rm max}=6$ behaves similarly to the
unconstrained case as far as liquid-gas phase separation is
concerned. We perform MD simulations to efficiently propagate the
system in time, even if Brownian dynamics (BD) would give a more
realistic description of the short time dynamics.  As far as slow
dynamics is concerned, MD and BD are
equivalent~\cite{gleim,ghost}.

We report in Fig.~\ref{fig:perc_spin} the evolution of the spinodal
line for different $n_{\rm max}$ in the $(\phi,T)$-plane.  We estimate
the spinodal line by bracketing it with the last stable state point
and the first phase separating state point along each isochore.  The
last stable state is characterized by a value of the structure factor
at low $q$
of the order of 10.  We confirm the location of the spinodal by
detecting the vanishing of the derivative of $P(V)$ along
isotherms. The unstable area in the $(\phi,T)$-plane shrinks on
decreasing $n_{\rm max}$, showing that the additional constraint opens
up a significant portion of phase space, where the system can be
studied in equilibrium one-phase conditions. The $\phi$ at which phase
separation is not present, at all $T$, can be as low as $\phi \approx
0.20$ for $n_{\rm max} = 3$ \cite{notasq}, $\phi\approx 0.30$ for
$n_{\rm max}=4$ and $\phi \approx 0.35$ for $n_{\rm max}=5$.
Fig.~\ref{fig:perc_spin} also shows the percolation lines, calculated
as the loci where $50\%$ of independent configurations are
characterized by the presence of a spanning cluster of bonded
particles.  The percolation loci also shift to lower $T$ on decreasing
$n_{\rm max}$, always ending in the spinodal on the low $\phi$ side.

As shown by the data in Fig.\ref{fig:perc_spin}, the addition of the
constraint on the maximum number of bonds, by suppressing the phase
separation, makes it possible to study the dynamics of the model at
very low $T$, where the lifetime of the interparticle bond increases,
stabilizing for longer and longer time intervals the percolating
network\cite{jack}.  When the bond lifetime becomes of the same order
as the observation time, the system will behave as a disordered solid.
It is worth stressing that in the present model there is no
thermodynamic transition associated with the onset of a gel
phase\cite{notagel}.  

To quantify these propositions we study in the following the dynamics
for all $T$ in the $\phi$
region where the system is in a single phase.

\begin{figure}[h]
\includegraphics[width=.45\textwidth]{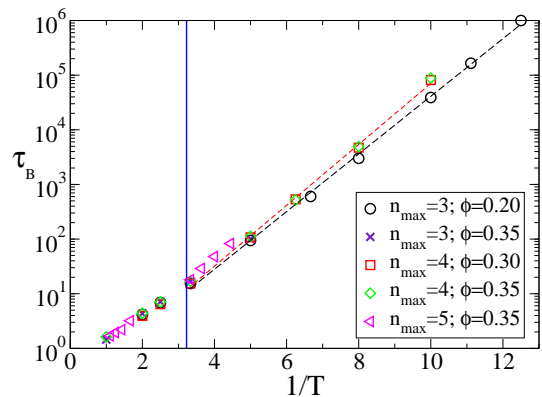}
\caption{
Arrhenius plot of the bond lifetime $\tau_B$ for different values of
$n_{\rm max}$ and $\phi$.  The vertical line indicates the lowest
accessible $T$ in the one-phase region for the unconstrained ($n_{\rm
max}=12$) case. Dashed lines are Arrhenius fits.
}
\label{fig:bond}
\end{figure}

Fig.~\ref{fig:bond} shows the $T$-dependence of the bond lifetime
$\tau_{B}$, defined as the time at which the bond autocorrelation
function decays to $0.1$, along different isochores for $n_{\rm
max}=3$, $4$ and $5$.  The bond lifetime follows an Arrhenius behavior
at low $T$, with an activation energy slightly increasing with $n_{\rm
max}$, of the order of the bond energy, suggesting that breaking of
one single bond is the elementary process.  
We do not observe any
significant $\phi$ dependence, suggesting that the bond-breaking
depends mostly on $T$\cite{speedybond} as well as on $n_{\rm
max}$. All curves, for all reported $n_{\rm max}$ and $\phi$,
superimpose onto each other for high $T$, recovering the HS limit. The
vertical line in Fig.~\ref{fig:bond} indicates the lowest $T$
which can be reached before encountering the coexistence region for
the unconstrained case. It is interesting to note that the
introduction of a small $n_{\rm max}$ makes it possible to explore
states with dynamics at least $4$ orders of magnitude slower than
without the constraint, allowing for an approach to arrested states
from equilibrium conditions.

\begin{figure}
\includegraphics[width=.39\textwidth]{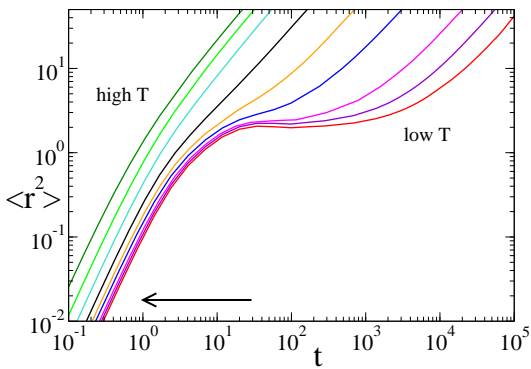}
\caption{
$\langle r^2\rangle$ for $n_{\rm max}=3$ and $\phi=0.20$ as a function
of $T$. From left to right, $T$s are: $1.0, 0.5, 0.3, 0.2, 0.15,
0.125, 0.1, 0.09, 0.08.$ The arrow signals the typical plateau value
for the HS glass.}
\label{fig:MSD}
\end{figure}

Fig.~\ref{fig:MSD} shows the $T$ dependence of the mean squared
displacement $\langle r^2\rangle$. A $T$-independent plateau develops
on cooling, indicating the presence of a localization length of order
$\sigma$, much larger than the corresponding value typical of glass
forming systems ( $\langle r^2\rangle \sim 0.01\sigma^2$). The
localization length decreases on increasing $\phi$. Caging in the gel
is thus induced neither by the bond length ($\Delta$) nor by the
nearest neighbor distance and hence it is significantly different
from the case of attractive and repulsive
glasses\cite{pham,natmat}. This feature signals the crucial role of
connectivity in the gel transition.

\begin{figure}
\includegraphics[width=.425\textwidth]{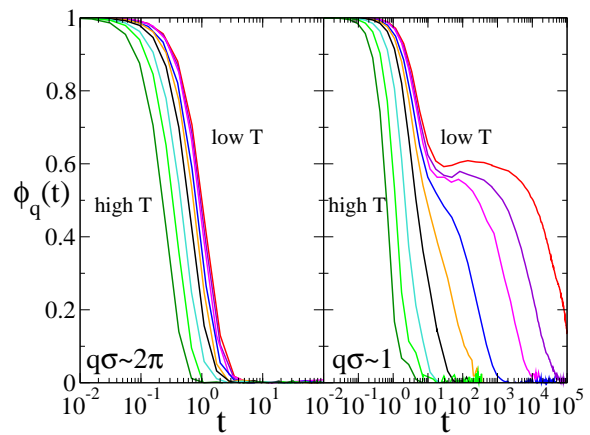}
\caption{
Density autocorrelation functions $\phi_q(t)$ for $n_{\rm max}=3$ and
$\phi=0.20$ as a function of $T$ (sequence of values as in
Fig.\protect\ref{fig:MSD}), respectively for $q\sigma \sim 2\pi$
(left) and $q\sigma \sim 1$ (right). Note the difference in $T$
dependence of $\tau_{\alpha}$ in the two cases.
}
\label{fig:fqt}
\end{figure}

To quantify the dynamical slowing down on approaching 
the gel transition, we study
the normalized density autocorrelation functions
$\phi_q(t)\equiv\left<\rho_q(t)\rho_{-q}(0)\right>/S(q)$.
Fig.\ref{fig:fqt} shows the behavior of $\phi_q(t)$ on decreasing
$T$. A striking dependence on $q$ is observed, a feature missing in
the slow dynamics close to the glass transition. If we focus, as
typically done in glass transition studies, on the $q$ value of the
first peak of $S(q)$, corresponding to the inverse average
nearest-neighbor distance, we observe no sign of a plateau in $\phi_q(t)$, 
within the precision of our calculations (left panel of Fig.~\ref{fig:fqt}). 
However, at smaller $q$ values, a clear plateau, 
named $f_q$, in analogy with its counterpart 
for the glass transition, where it is usually named non-ergodicity factor,
emerges at a $T$ that varies with $q$. 
This behavior is profoundly different from what is observed in a standard glass
transition, but closely resembles what is observed at a percolation
transition in the presence of chemical (infinite lifetime) bonds
\cite{delgado,barrier}. 
We find that the $T$-dependence of the relaxation time $\tau_{\alpha}$
(defined as the time at which $\phi_q(t)$ reaches the value $0.1$)
crosses over from $\tau_{\alpha}\sim 1/\sqrt{T}$ at nearest neighbour
distances to $\tau_{\alpha}\sim e^{1/T}$ at large length scales. For
small $q$, $\tau_{\alpha}$ is coupled to the bond lifetime, while at
larger $q$, $\tau_{\alpha}$ is controlled by the microscopic time
scale (proportional to the thermal velocity).
\begin{figure}
\includegraphics[width=.425\textwidth]{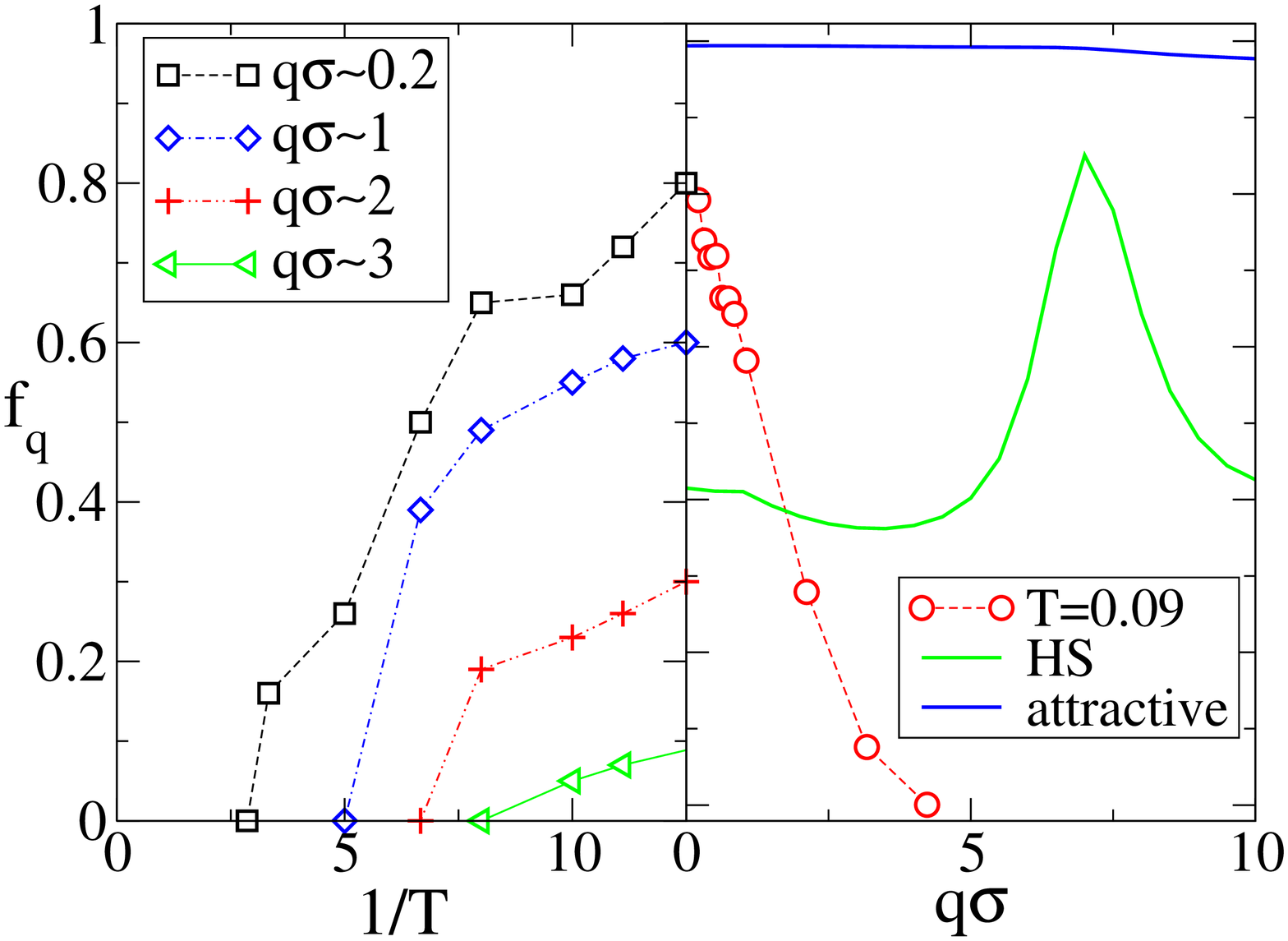}
\caption{Left panel: Plateau height $f_q$ for $n_{\rm max}=3$ and 
$\phi=0.20$ as a function of $T$, for different values of $q\sigma$.
At a given $T$, $f_q\neq 0$
at small $q$.  At larger $q$, though $q\sigma$ is still smaller than
$2\pi$ (nearest neighbor length scale), $f_q=0$ 
within the accuracy of the data. Right panel: $f_q$ at $T=0.09$,
$\phi=0.20$, $n_{\rm max}=3$, compared with $f_q$ of attractive and HS
glass \protect\cite{fqpy}.  }
\label{fig:fq}
\end{figure}
Fig.~\ref{fig:fq} shows the $T$-dependence of the plateau value $f_q$
(calculated as the amplitude of a stretched exponential fit to the
slow relaxation) for various $q$'s (left panel), and compares its $q$
dependence with $f_q$'s typical of HS and attractive
glasses\cite{fqpy} (right panel).  While in the case of glasses
(either HS or attractive) $f_q$ is significantly different from zero
at all physically relevant $q$ values, in the present case $f_q$ is
significantly different from zero only at very small wavevectors.  It
is also interesting to observe that, while in the case of glasses
$f_q$ jumps from zero to a finite value at the glass transition, in
the present case $f_q$ appears to grow smoothly from zero.

In summary, we have proposed an off-lattice model which allows the
study of dynamical arrest at low $\phi$ (physical gelation) in the
absence of macro or micro phase separation, effectively decoupling the
effects of phase separation from the dynamical slowing down. The
process of arrest
can be followed in equilibrium. This condition offers the possibility
of studying the behavior of the density correlators and of the mean
squared displacement close to the gel transition, which is strongly
coupled to the bond dynamics.  We discover significant differences
between glasses and gels in terms of the decay of the density
correlation functions and of the localization length, which could be
studied experimentally to discriminate between the phase separation
route and the {\it single phase} reversible gelation case.

We acknowledge support from MIUR Cofin and Firb, MRTN-CT-2003-504712,
NSERC-Canada (I.~S.-V.), NSF (S.~V.~B.)  and DIPC-Spain (A.~J.~M.).
We thank E. Del Gado for discussions and for sharing her unpublished
results and A. Puertas, R. Schilling and N. Wagner for comments.


\begin{thebibliography}{999}
\bibitem{notagel}
The important issue of the thermodynamic nature of the gel transition
has been recently reviewed in Ref.\protect\cite{rubinstein}.  In our
model, the free energy can be calculated numerically with high
precision from high $T$, where the system reduces to a HS fluid, down
to very small $T$, where particles are fully connected into a single
cluster. Indeed, the finite bond lifetime guarantees equilibrium even
well within the geometric percolation region.  The free energy and its
derivatives are continuous in the entire $T$-range, proving that
across the gelation transition there is no thermodynamic transition.
A locus of specific heat extrema, arising from the two-state origin of
particle bonding, is present in the phase diagram around $T/u_0
\approx 0.25$. Our model strongly
supports the view that gelation is a connectivity transition (see
Fig. \protect\ref{fig:MSD}) with no thermodynamic character. 
Indeed, the  cluster size distribution at the percolation transition is
consistent
with random bond percolation. We find  that the cluster size
distribution $n_s \sim s^{-\tau}$, with $\tau=2.20 \pm 0.05$, to be
compared with the theoretical
value $\tau=2.18$.
\end{thebibliography}
\end{document}